# Interface observation of heat-treated Co/Mo$_2$C multilayers

Yanyan Yuan[1,2], Karine Le Guen[1,2], Jean-Michel André[1,2], Christian Mény[3], Corinne Ulhaq[3], Anouk Galtayries[4], Jingtao Zhu[5], Zhanshan Wang[5], Philippe Jonnard[1,2*]

[1] *Sorbonne Universités, UPMC Univ Paris 06, Laboratoire de Chimie Physique-Matière et Rayonnement, 11 rue Pierre et Marie Curie, F-75231 Paris cedex 05, France*

2 *CNRS UMR 7614, Laboratoire de Chimie Physique-Matière et Rayonnement, 11 rue Pierre et Marie Curie, F-75231 Paris cedex 05, France*

[3] *Institut de Physique et Chimie des Matériaux de Strasbourg, UMR 7504 CNRS-Université De Strasbourg 23 rue du Loess, 67034 Strasbourg, France*

[4] *PSL, Research University, Institut de Recherche de Chimie Paris, CNRS – Chimie ParisTech, 11 rue Pierre et Marie Curie, F-75005 Paris, France*

[5]*Institute of Precision Optical Engineering, Department of Physics, Tongji University, Shanghai 200092, China*

**Abstract**

We study the interface evolution of a series of periodic Co/Mo$_2$C multilayers as a function of the annealing temperature up to 600°C. Different complementary techniques are implemented to get information on the phenomenon taking place at the interfaces of the stack. The periodical structure of Co/Mo$_2$C multilayer is proven by Time-of-flight secondary ion mass spectrometry (ToF-SIMS) depth profiles which demonstrate the formation of an oxide layer at both air/stack and stack/substrate interfaces. From Nuclear magnetic resonance (NMR) spectra, we observed the intermixing phenomenon of Co and C atoms for the as-deposited sample, and then at annealing temperature above 300°C Co and C atoms separate from their mixed regions. Comparison of NMR results between Co/Mo$_2$C and Co/C references confirms this phenomenon. This is in agreement with x-ray emission spectroscopy (XES) measurements. Furthermore the calculation of the Co-C, Co-Mo and Mo-C mixing enthalpy using Miedema's model gives a proof of the demixing of Co and C atoms present within the stacks above 300°C. From the transmission electron microscopy (TEM) analysis, we found the presence of some crystallites within the as-deposited sample as well as the mainly amorphous nature of all layers. This is confirmed using x-ray diffraction (XRD) patterns which also demonstrate the growth of crystallites induced upon annealing.

*Corresponding author: E-mail: philippe.jonnard@upmc.fr. Tel: +33 01 44 27 63 03, Fax:+33 01 44 27 62 26.

# 1. Introduction

Multilayer systems periodically alternating (at least) two materials are widely used in the extreme ultra-violet and soft x-ray ranges. The optical performances of multilayers mirrors are usually limited by the interface and surface quality, such as intermixing, interdiffusion, interface reaction and surface oxidation. Multilayer optics may be subject to high temperatures in their applications, for instance intense synchrotron light, free electron laser or laser plasma sources. Thus the ability to withstand high-power x-ray radiation without undergoing radiation damage or interdiffusion is an important merit for an x-ray multilayer mirror.

A lot of works focused on the study of the thermal behaviour of the Mo/Si [1–5], Al-based [6–9] and Co-based multilayers [10–14] have demonstrated that interface properties play an important role in the optical performances. Concerning these multilayers, single elements are usually used as a spacer and absorber layers to constitute a multilayer. However it is generally easy to form some compounds between two single elements. A possible solution to avoid or at least reduce the interdiffusion and compound formation is to build a multilayer where one or even both spacer and absorber layers consist in a stable compound. Bai *et al.* reported that CoMoN/CN multilayers show relatively stable interfaces [15]. Recently, the investigation of $B_4C/Mo_2C$ multilayer also demonstrated abrupt interfaces with respect to $B_4C/Mo$ multilayer where interlayers were detected [16].

The evolution of interface structure upon annealing has been widely studied. In Mo/Si multilayers, it has been reported that the reflectivity decreases upon annealing owing to the increase of interdiffusion at interfaces [17]. The Co/C multilayers showed an enhanced reflectivity at annealing temperatures lower than 250°C due to phase separation [18]. Moreover a backward diffusion of two elements, from their mixtures to the pure elements, at the interfaces was also observed in the Co/Au, Co/Ag, Co/C and CoN/CN multilayers annealed up to 250°C [19–22].

In the present paper, we study the change of the interface properties of the Co/Mo$_2$C multilayer with the annealing temperature. The optical properties of this series multilayers have been investigated by using hard x-ray and soft x-ray reflectivities and already reported in Ref. [23]. To summarize, the best reflectivity is 27% at 1.59 nm for 300°C annealed sample and decreases down to 20% following annealing at 600°C. The interface structure is studied by combining both non-destructive methods, x-ray emission spectroscopy (XES), nuclear magnetic resonance (NMR) spectroscopy, x-ray diffraction (XRD) and destructive methods, time-of-flight secondary ions mass spectroscopy (ToF-SIMS) and transmission electron microscopy (TEM). The comparison of the XES Co/Mo$_2$C multilayer spectra with that of the reference samples, a Mo$_2$C thin film and Co/C multilayer, is used to determine the chemical state of the C atoms within the stack. NMR spectroscopy is carried out to provide the information about the chemical state of Co atoms by describing the Co atoms distribution as a function of their resonance frequency. ToF-SIMS depth profiles determine the distribution in depth of various elements. Phase change and microstructure are studied by using XRD and transmission electron microscopy (TEM) respectively.

## 2. Experiments

All the samples are prepared following the designed parameters described in a previous paper[24]. Let us recall that the samples were prepared on the sliced and polished Si(100) substrate by magnetron sputtering. The based pressure was $10^{-5}$ Pa before deposition. The sputtering gas was argon (99.999% purity) at a constant working pressure 0.1 Pa. The powers applied to sputtering target were 40 W and 60 W for Co and Mo$_2$C, respectively. The aimed period is equal to 4.1 nm and the ratio of the Mo$_2$C thickness to the period is 0.36, resulting in Co and Mo$_2$C thicknesses of 2.6 nm and 1.5 nm, respectively. Six samples with an area of 20 x 20 mm$^2$ were deposited separately on the Si wafer with the same conditions. From one sample to another a small variation of the period can exist. The number of periods is 30. A 3.5 nm-thick B$_4$C capping layer was deposited on top of the samples to prevent oxidation. The samples were annealed at 200, 300, 400, 500 and 600°C for 1 h under high vacuum. A reference $Mo_2C$ thin film, with a thickness of about 60 nm, was deposited onto a silicon substrate with the same deposition conditions as the $Mo_2C$ layers of the multilayer. Three Co/C multilayers with the thickness of 2.0 nm for both Co and C layers in one period were deposited on a silicon substrate using the same conditions as for the Co/$Mo_2C$ multilayers. Two of them were annealed at 300°C and 600°C for one hour in a furnace with a base of pressure of $3.0\times10^{-4}$ Pa.

## 2.1. X-ray emission spectroscopy

The X-ray emission analysis was performed in a high-resolution wavelength dispersive soft x-ray spectrometer [25] equipped with a curved etched multilayer [26] to disperse the emitted radiation. The C K$\alpha$ emission (2p – 1s transition) from the Co/$Mo_2C$ multilayers and the reference samples ($Mo_2C$ thin film and Co/C multilayer) was analysed. Graphite was also used as a reference material. The Mo valence states have not been studied since the corresponding emission (Mo L$\beta_2$, 4d – 2p$_{3/2}$ transition), accessible with our spectrometer, is not very sensitive to the chemical state of the molybdenum atoms. The same is true with the Co atoms [27].

## 2.2. Nuclear magnetic resonance spectroscopy

In order to probe the chemical state of the Co atoms within the multilayer, the samples were analyzed by zero-field NMR spectroscopy. The NMR spectra represent the distribution of the Co atoms as a function of their resonance frequency. The NMR resonance frequency is sensitive to the local environment of the probed atoms: the number, nature and symmetry of atoms in its neighbourhood [28,29]. The NMR spectroscopy of a $Co_3Mo$ disordered alloy and two Co/C multilayers (as-deposited and 600°C) were also measured as reference samples to analyse the Co/$Mo_2C$ spectra. To enhance the sensitivity, the measurement temperature was 2 K for all the multilayer samples and 4.2 K for the $Co_3Mo$ disordered alloy. All the multilayer spectra are normalized to the samples surface area.

### 2.3. Time-of-flight secondary ions mass spectrometry

The depth distribution of the Co/Mo$_2$C multilayers was investigated using a ToF-SIMS instrument (TOF.SIMS V spectrometer, ION-TOF GmbH) working in the dual-beam mode. The sputtering was performed using a 1 keV (59 nA) Cs$^+$ ion beam, rastered over an area of 300 μm × 300 μm. A pulsed 25 keV Bi$^+$ primary ion source at a current of 1.3 pA (high current bunched mode), rastered over a scan area of 100 μm × 100 μm was used as the analysis beam. Both ion beams were impinging the sample surface forming an angle of 45° with the surface normal and were aligned in such a way that the analyzed ions were taken from the center of the sputtered crater. Both positive and negative ions were recorded. In the present paper, only negative ions spectra are presented since they can provide sufficient information from each layer. The ToF-SIMS technique provides a qualitative chemical analysis as the intensity does not correspond directly to the number of atoms in the multilayer (the sputtering yields vary from one element to another as well as from one matrix to another).

### 2.4. X-ray diffraction

To further characterize the phase structure of samples annealed at different temperatures, the XRD measurements were performed by using a PANalytical X'Pert Pro diffractometer with Cu Kα radiation (0.154 nm) emitted from an x-ray tube operating at 45 kV and 40 mA. The scanning angle range is 10-80° with a step of 0.01°. For the sake of comparison, the XRD patterns of the Co/C reference multilayers were also performed. Here no refinement of procedure of the structural characterization [30, 31] was done.

### 2.5. Transmission electron microscopy

Ultra-thin cross-sectional slices were prepared using a focused ion beam (FIB). High-resolution images were obtained by the JEOL 2100F instrument operated at 200 kV with an image resolution of 0.2 nm. Scanning transmission electron microscopy (STEM) images were obtained by scanning the focused beam over the slice with an image resolution of 0.11 nm. High angle-annular dark field (HAADF) images can be collected simultaneously, which makes the STEM a good technique for materials morphological and chemical characterization of materials at the sub-nanometer level.

## 3. Results

### 3.1. X-ray emission spectroscopy

The C K$\alpha$ emission bands emitted by the as-deposited and annealed at 600°C Co/Mo$_2$C multilayers as well as the Mo$_2$C, graphite and Co/C multilayers references (for this latter, as-deposited and annealed at 600°C), are presented in Fig.1. The spectrum of the Co/Mo$_2$C annealed at 300°C is the same as that of the as-deposited sample and is thus not shown here. The spectra of the Co/Mo$_2$C annealed at 400 and 500°C are not shown here as they are the same as that of Co/Mo$_2$C annealed at 600°C (Fig.1(a)).

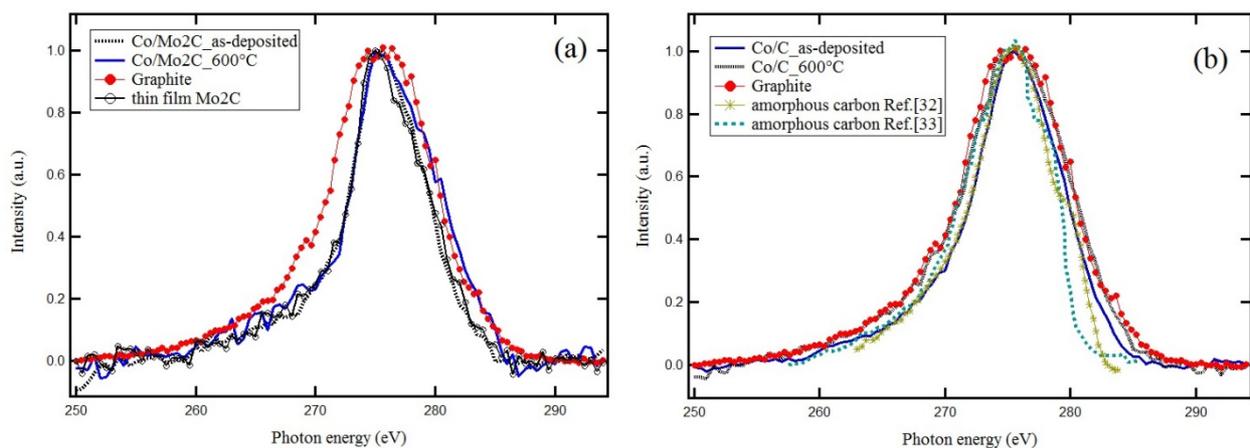

Fig.1. C K$\alpha$ emission bands emitted by the as-deposited and annealed at 600°C Co/Mo$_2$C multilayers as well as the thin film Mo$_2$C, graphite (a) and by (as-deposited and annealed at 600°C) Co/C references, graphite and amorphous carbon from Refs.[32,33] (b).

Regarding the Co/C reference multilayers, the spectrum of the as-deposited Co/C sample should be similar to one of the amorphous carbons if no interaction takes place within the structure. The comparison with the spectra of amorphous C from Refs. [32,33] shows that this is not the case (Fig.1(b)). This indicates that, in the as-deposited Co/C system, some interaction exists between the C and Co atoms leading to intermixing between C and Co layers. The consequence is a change in the density of C (and Co) valence states and thus in the shape of the C K$\alpha$ emission band. With respect to as-deposited Co/C, the spectrum of Co/C annealed at 600°C shows a significant broadening and is close to that of graphite. These results suggest that the Co and C atoms are intermixed within the as-deposited Co/C, while the annealing at 600°C induces the demixing of Co and C atoms and the "graphitization" of C layers.

The spectrum of as-deposited Co/Mo$_2$C is the same as that of both as-deposited Co/C and of the Mo$_2$C references. Thus the chemical state of the C atoms is not distinguishable for the Co/Mo$_2$C as-deposited multilayer: the carbon atoms exist either in the form of Co and C mixture or in the form of Mo$_2$C compound or in both forms. Following annealing at 400°C and above, the spectra of the Co/Mo$_2$C multilayers show a slight broadening towards the high photon energy side. This suggests that a slightly change of the chemical state of the C atoms takes place from 400°C.

3.2. Nuclear magnetic resonance spectroscopy

NMR spectra of all Co/Mo$_2$C multilayers as a function of the annealing temperature are presented in Fig.2. It can be seen that the spectra of the as-deposited and 200°C samples consist of a wide structure without any extra feature. For the 300, 400, 500 and 600°C annealed samples, a well-defined line is observed at around 221 MHz. This line corresponds to bulk Co with an *hcp*-like structure [34]. However the frequency is lower than the one expected for well crystallized *hcp* Co. As suggested in Ref. [35], this is most probably due to the presence of a large amount of stacking faults within the layer. Regarding the as-deposited and 200°C annealed samples the absence of a well-defined NMR feature indicates that most of Co atoms are mixed with other atoms (Mo or C). From 300°C, the intensity of the 221 MHz line increases with the annealing temperature while the low frequency (<220 MHz) contributions stay almost unchanged. This indicates that the total amount of ferromagnetic Co atoms as well as pure Co increase from 300°C and above. From the phase diagram of Co-Mo system showing the existence of several ordered phases, it is unlikely that annealing the samples result in a phase separation between Co and Mo. Therefore the strong intermixing revealed by the as-deposited NMR spectra is more likely to come from mixing of Co and C atoms. Since the Co carbides are not stable phases the C is likely to separate from Co-C mixing region upon annealing. This phenomenon has also been reported by H. Bai *et al.*[21].

To check this interpretation, we measured the NMR spectrum of a reference $Co_3Mo$ disordered alloy, as well as the spectra of the Co/C multilayers as-deposited and annealed at 600°C, Fig.3. The spectra of the $Co_3Mo$ disordered alloy and of the Co/C as-deposited multilayer show a low frequency signal (<220 MHz) without bulk Co peak. However in both cases the detailed shape of the NMR spectra is different from that of the $Co/Mo_2C$ as-deposited sample. More interestingly, the spectrum of the Co/C sample annealed at 600°C shows an intense peak at about 220 MHz that can be attributed to *hcp-like* Co. This behaviour is very similar to the one of the $Co/Mo_2C$ annealed multilayers. Therefore this strongly suggests that the phase separation observed after annealing of the $Co/Mo_2C$ samples is due to the incorporation of C onto the Co layers of the as-deposited samples, the molybdenum staying outside the Co layer. The difference between the $Co/Mo_2C$ and the Co/C multilayers is attributed to the fact that the carbon content in the Co layer is different in the Co/C and in the $Co/Mo_2C$ multilayers.

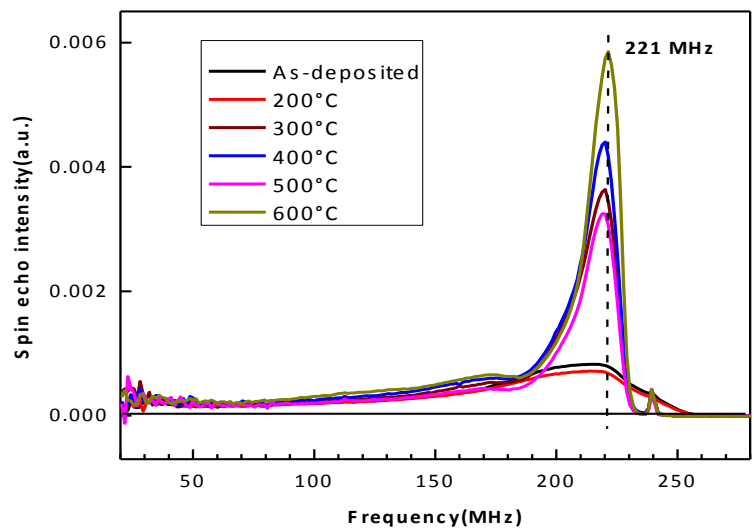

Fig.2 NMR spectra of the $Co/Mo_2C$ multilayers as a function of the annealing temperature.

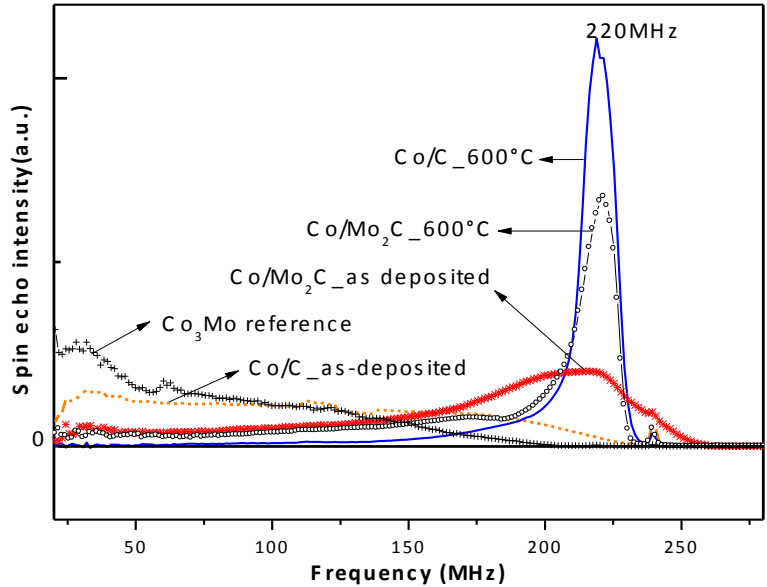

Fig.3 NMR spectra of as-deposited and annealed at 600°C Co/Mo$_2$C, compared to the NMR spectra of Co$_3$Mo disordered alloy as well as as-deposited and annealed at 600°C Co/C references.

3.3. X-ray diffraction

The x-ray diffraction patterns of the as-deposited and 200°C to 600°C annealed Co/Mo$_2$C samples are presented in Fig. 4(a). A broad and low intensity Bragg peak is observed at around 44° which corresponds to the Co (111) or Co$_3$Mo (002) phases. The intensity of this feature increases slightly from the as-deposited sample to the 500°C annealed sample. After annealing at 600°C, the intensity increases much more showing a significant structural change. This phenomenon can be related to the recrystallization of the Co layers owing to the demixing of Co and C atoms after annealing or the crystallization of the disordered Co$_3$Mo mixture after annealing at high temperature.

In order to explain the change between Co and C layers upon annealing, we performed the XRD patterns of the Co/C multilayers (Fig. 4(b)). There is no significant peak to be seen. This means that the Co/C multilayers are mainly amorphous and the Co layers are not crystallized. It should be noted that the thickness of Co layer in one period is about 2 nm. So the probability of forming a crystalline phase in the stack is low.

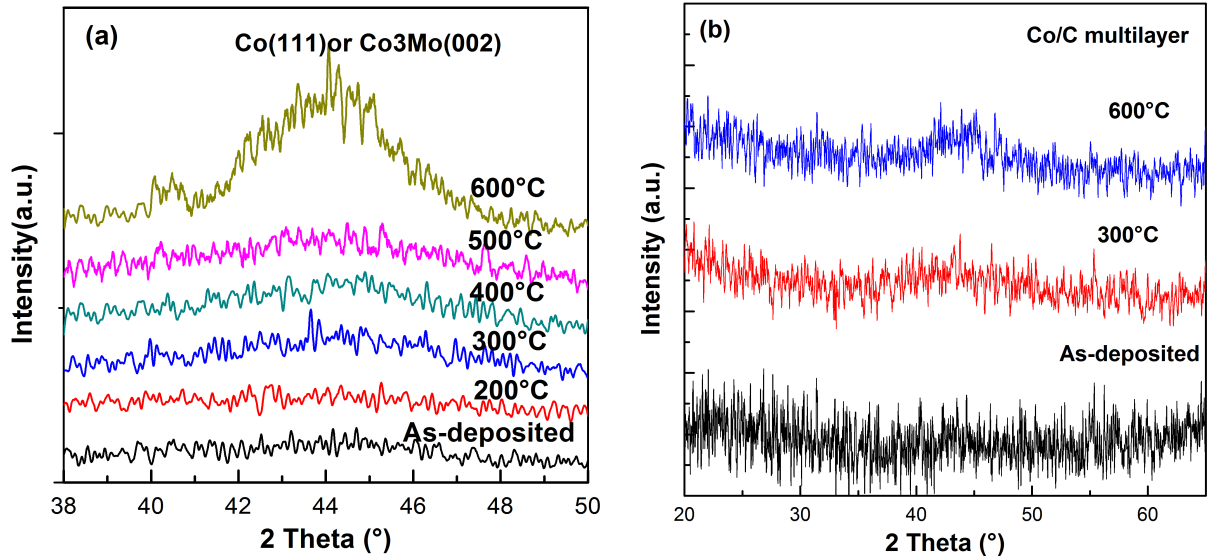

Fig.4 Diffraction patterns of the Co/Mo$_2$C samples: as-deposited, annealed at 200°C, 300°C, 400°C, 500°C and 600°C (a) and of the Co/C multilayers: as-deposited, 300 and 600°C (b).

3.4. Time of flight-secondary ions mass spectroscopy

Shown in Fig.5 are the ToF-SIMS depth profiles of C$^-$ ions of the as-deposited, 300 and 600°C Co/Mo$_2$C samples. The amplitude of the signal in the middle of the profile is normalized to unity to show the relative contrast of the ion signal. Then, for the sake of clarity, the curves of 300°C and 600°C annealed samples are arbitrarily shifted. The periodic modulation of C$^-$ ions signal gives evidence of the periodic structure of the multilayer. There are only 29 periods for these three samples from the outermost period to the substrate, meaning that the 30$^{th}$ period is chemically modified. At the stack/substrate interface, this period is mixed with the Si substrate and the C$^-$ profile presents a low amplitude signal at the end of the profile.

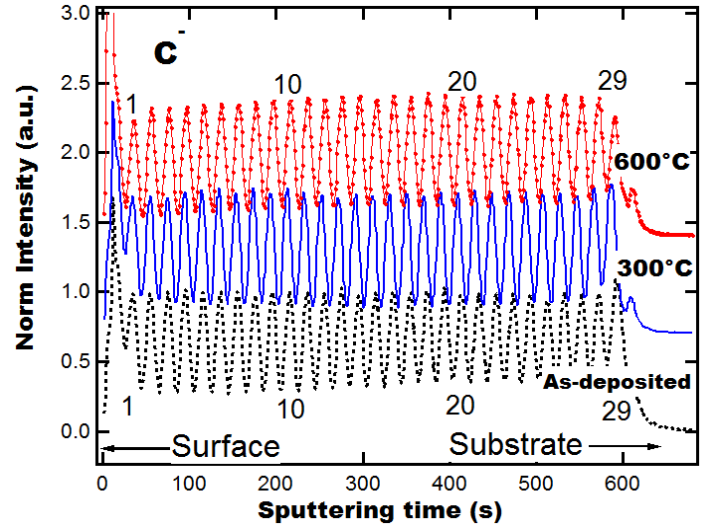

Fig.5 ToF-SIMS depth profiles of the C⁻ ions for the as-deposited, 300°C and 600°C annealed Co/Mo$_2$C samples(Cs⁺, 1 keV, 59 nA).

The C⁻, Co⁻, O⁻, CoO⁻, B⁻ and MoO$_2$⁻ profiles toward the air/stack and stack/substrate interfaces for the as-deposited, 300 and 600°C annealed samples are presented in Fig. 6. From the O⁻ depth profiles, it can be deduced that oxidation took place at both interfaces. Oxidation at the air/stack interface is due to the atmospheric contamination of the surface layers during storage in air and transfer. Oxidation at the stack/substrate interface is due to the native oxide on the silicon substrate. Similar phenomena were reported in Al/Mo/SiC multilayers [36]. The oxidation of the first Mo$_2$C layer in the stack was also reported by Giglia *et al.*[37]. At the stack/substrate interface, the O⁻ profile for the 300°C and 600°C annealed samples is sharper than that of the as-deposited sample. This is probably caused by the diffusion of oxygen atoms from the substrate into the stack after annealing. We did not get the information about the combination between O and Mo atoms as no difference in the MoO$_2$⁻ profile can be found when the annealing temperature is varied. The oxidation at the stack/substrate interface explains why the 30[th] period is not well detected.

Furthermore, in the case of as-deposited and 300°C annealed samples, the signal of the CoO⁻ ions at the air/stack and stack/substrate interfaces is more intense than in that of the 600°C annealed sample. For 600°C annealed sample, the signal of the CoO⁻ ions is not detected at the air/stack interface. Firstly, this indicates that Co layers are oxidized at temperatures lower than 300°C and that the Co oxide decomposes at high temperature. The C⁻ and B⁻ profiles do not show any significant change in the three samples.

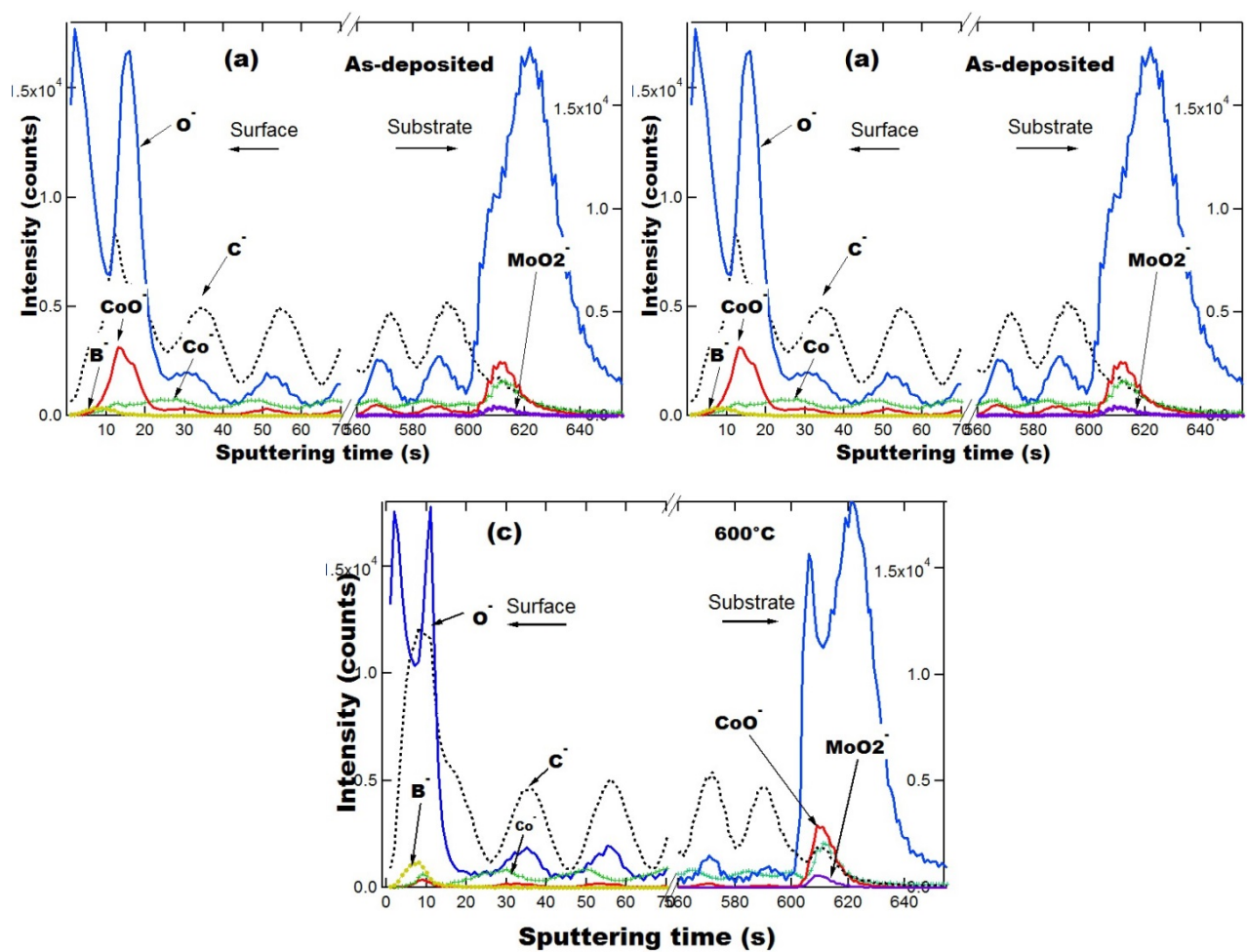

Fig.6 ToF-SIMS depth profiles of the C⁻, Co⁻, O⁻, CoO⁻, B⁻ and MoO₂⁻ ions at the air/stack and stack/substrate interfaces domains for as-deposited (a), 300 (b) and 600°C (c) annealed samples. (Colour online)

Figure 7(a) shows the ToF-SIMS depth profiles of the Co⁻, CoMo⁻ and C⁻ ions of the 600°C annealed sample. The Mo⁻ profile follows the C⁻ profile (not shown). The C⁻ profile displays a symmetrical shape and is almost in the opposite phase with respect to the Co⁻ profile. The Co⁻ depth profile displays some shoulders which correspond to the maxima in the CoMo⁻ depth profile. This is attributed to a matrix effect since for ultra-thin layers it is not possible to reach a steady sputtering regime before the end of the period. Figure 7(b) shows the comparison of the Co⁻ and C⁻ ions profiles of the as-deposited and 600°C annealed samples. Both ions profiles are normalized. We cannot observe any significant changes in the Co⁻ depth profile between as-deposited and 600°C annealed samples. The same situation is recorded for the C⁻ depth profile except a slightly better contrast for the sample annealed at 600°C with respect to the as-deposited sample.

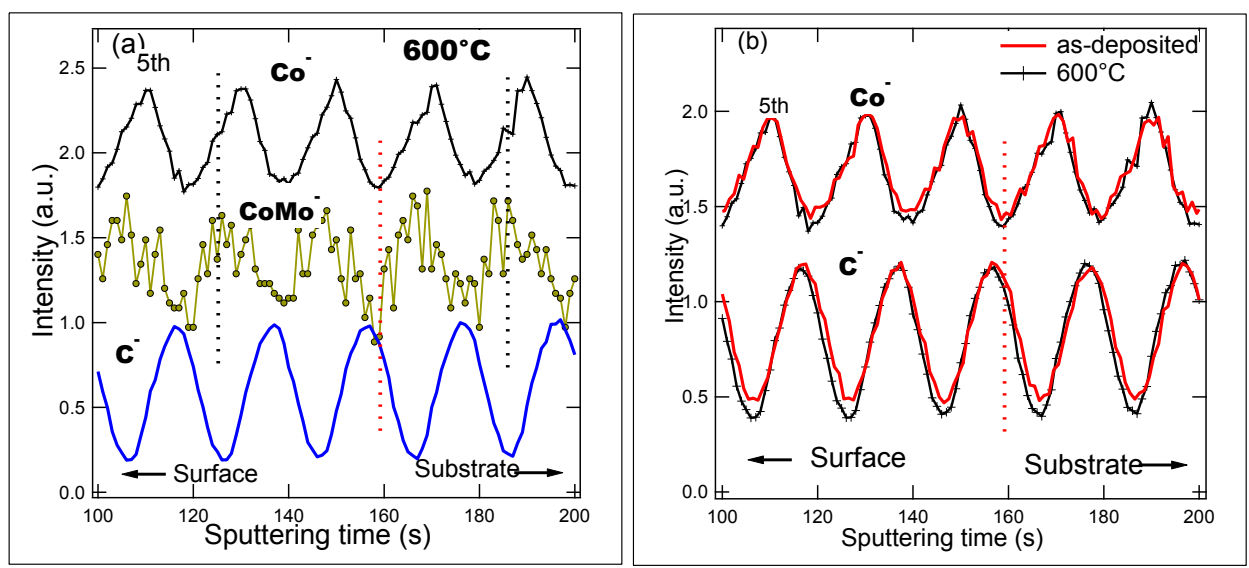

Fig.7 ToF-SIMS depth profiles of the Co⁻, CoMo⁻ and C⁻ ions for the 600°C annealed sample (a) and comparison of depth profiles of the Co⁻ and C⁻ ions for the as-deposited and 600°C annealed samples (b). ToF-SIMS depth profiles of Co⁻ and CoMo⁻ ions are shifted vertically for clarity.

3.5. Transmission electron microscopy

Figure 8(a) and (b) show the HAADF and bright-field images of the Co/Mo$_2$C as-deposited sample, respectively. The HAADF image shows a nice periodic multilayer structure since the individual layers are smooth. In the bright field image, Co layers appear bright and Mo$_2$C layers appear dark. We observe the appearance of dark areas in some Co layers, which suggests the presence of some crystallites. In the high-resolution image shown in Fig.8(c), we can observe the lattice fringes within the Co layers and at the interfaces, even if the contrast between Co and Mo$_2$C layers is poor, owing to their close mean atomic number. Selected-area diffraction, shown in Fig 8(d), gives evidence of mainly amorphous structure in the stack. This observation is in agreement with the results of XRD. It is noted that crystallites cannot be detected in the XRD techniques.

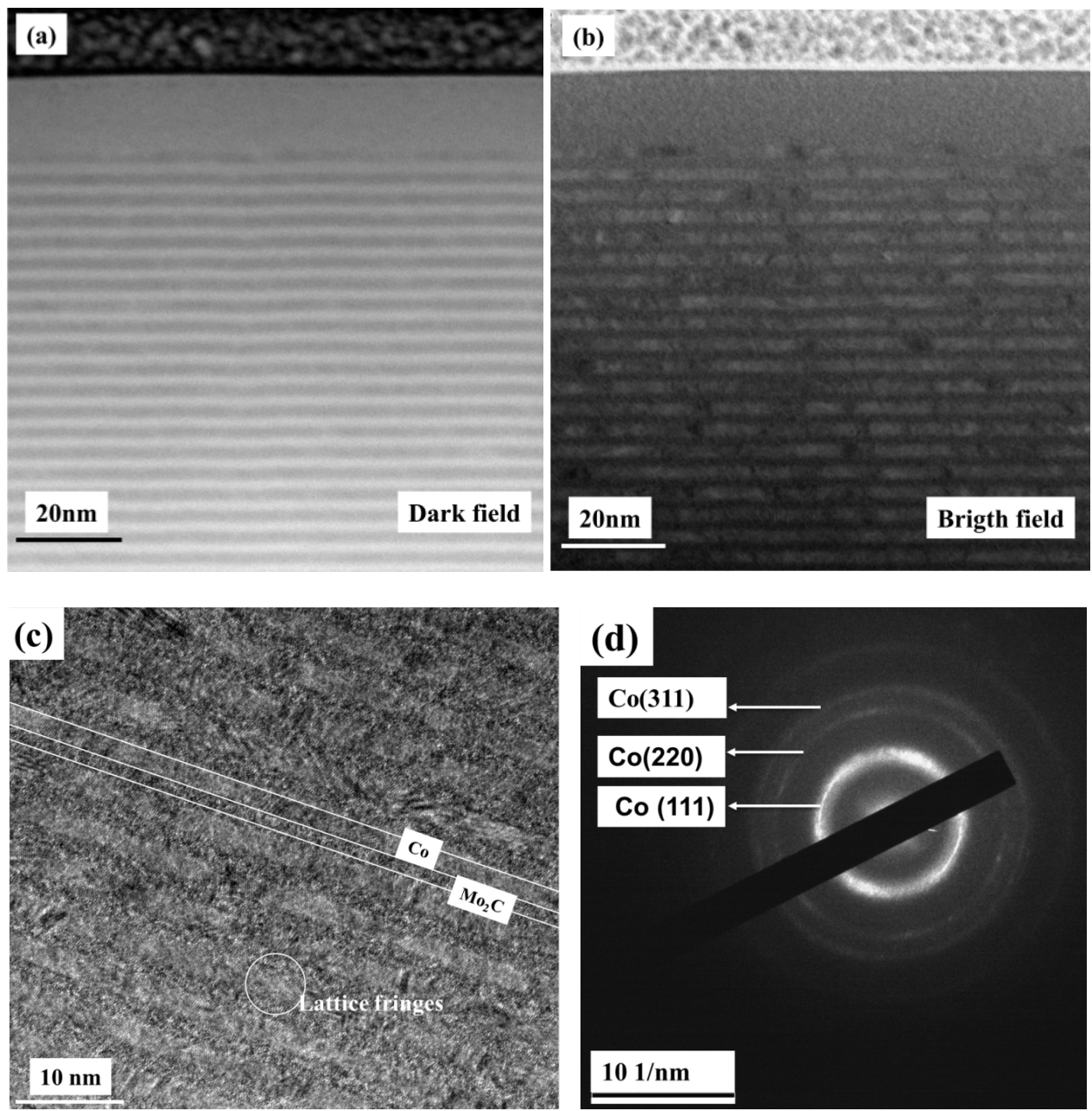

Fig 8 Transmission electron microscopy of the Co/Mo$_2$C as-deposited sample: HAADF (a) and bright field image (b); high-resolution image (c) and selected-area electron diffraction (d).

## 4. Discussion

From the preceding analysis, it suggests that the enhancement of ferromagnetic Co signal above 300°C should be primary ascribed to a separation of Co and C atoms at the interfaces. In the Co-C binary phase diagram, there is no stoichiometric compound resulting from the reaction between Co and C [38]. In the Co-Mo binary phase diagram, Co$_3$Mo and Co$_7$Mo$_6$ (μ and δ phase) are mentioned [39,40]. The Co-C-Mo ternary phase diagram gives two ternary compounds

Mo$_3$Co$_3$C and Mo$_6$Co$_6$C but there is a small possibility for their formation in the Co/Mo$_2$C multilayer because of their high crystallization temperature [41]. Otherwise, the XRD results show that no significant Co$_3$Mo crystallites can be formed from as-deposited to 500°C annealed samples. Therefore, we can deduce that intermixing takes place for the as-deposited and 200°C annealed samples without significant formation of Co$_3$Mo compound.

Based on the Miedema's macroscopic atom model [42], we can calculate the mixing enthalpy of the Co-Mo and Co-C systems to confirm our results. According to this model, which have been developed to calculate the enthalpy values for binary alloys, the enthalpy of formation of the Co-Mo system, made of two transition metals, can be expressed as:

$$\Delta H = \frac{2Pc_{Mo} f_{Co}^{Mo} V_{Mo\,alloy}^{2/3} \left[ -(\phi_{Mo} - \phi_{Co})^2 + \frac{Q}{P}\left( n_{ws}^{Mo\,1/3} - n_{ws}^{Co\,1/3} \right)^2 \right]}{n_{ws}^{Mo\,-1/3} + n_{ws}^{Co\,-1/3}} \quad (1)$$

where

$$f_{Co}^{Mo} = C_{Co}^{S} \qquad \text{for non-ordered alloy}$$

$$f_{Co}^{Mo} = C_{Co}^{S}\left(1 + 8\left(C_{Mo}^{S} C_{Co}^{S}\right)^2\right) \qquad \text{for ordered alloy}$$

$$c_{Mo}^{S} = c_{Mo} V_{Mo}^{2/3} \left( c_{Mo} V_{Mo}^{2/3} + c_{Co} V_{Co}^{2/3} \right)$$

$$c_{Co}^{S} = c_{Co} V_{Co}^{2/3} \left( c_{Mo} V_{Mo}^{2/3} + c_{Co} V_{Co}^{2/3} \right)$$

$$P = 14.1, \qquad Q/P = 9.4 \; V^2/(d.u.)^{2/3}$$

Here, $c_{Co}$ and $c_{Mo}$ are the mole fraction ($c_{Co} + c_{Mo} = 1$), $V_{Co}$ and $V_{Mo}$ are the atomic volumes of Co and Mo, respectively, $\phi$ the electronegativity and $n_{ws}$ the electron density at the first Wigner-Seitz boundary. The electron densities are usually given in terms of density units (d.u.), 1 d.u.=6.75×10$^{22}$ electron/cm$^3$. $C_{Co}^{S}$ and $C_{Mo}^{S}$ are the surface densities of Co and Mo atoms, respectively. $f_{Co}^{Mo}$ represents the degree to which an atomic cell of Mo is in contact with dissimilar atomic cell of Co on average within the alloy. When we calculate the atomic volume, the atomic volumes of Co and Mo atoms are approximate the same ones. In equation (1), the first term is negative and proportional to the square of difference in the electronegativity parameters and therefore the charge transfer between Co and Mo. This part shows the tendency for compound formation. The second term which is positive stands for the discontinuity in electron density at the Wigner-Seitz boundary of Co and Mo atoms. This part shows the tendency of phase separation.

Another effect, which has to be considered, is the volume change of the constituents Co and Mo that occurs due to charge transfer when the alloy is formed. The charge transfer occurring between dissimilar atomic cells depends on the difference of the electronegative parameters of Co and Mo. An empirical relationship for volume change has been established

$$V_{Mo\,alloy}^{2/3} = V_{Mo\,pure}^{2/3} \cdot \left[1 + af_{Co}^{Mo}(\phi_{Mo} - \phi_{Co})\right] \quad (2)$$

where $a$ is a constant equal to 0.04 for Mo and 0.1 for Co [43].

Table 1. Values of $\phi$, $n_{ws}$ and $V$ for calculating the mixing enthalpy of the Co-Mo system.

| Element | $\phi$ (V) | $n_{ws}$ (d.u.) | $V$ (cm$^3$/mol) |
|---|---|---|---|
| Co | 5.10 | 7.00 | 6.7 |
| Mo | 4.37 | 8.46 | 9.4 |

Table 1 lists the values of $\phi$, $n_{ws}$ and $V$ used for calculating the mixing enthalpy ΔH. The calculated enthalpy of Co-Mo mixing system is shown in Fig.9. It is always negative. This indicates that the Co and Mo compounds easily form. Moreover the mixing enthalpy is negative when the mole fraction of Mo changes from 0 to 0.45. However the ratio of atomic fraction between Co and Mo in the Co/Mo$_2$C multilayer stack is about 3:1 and thus the Mo atomic fraction is 0.25 (according to the designed thickness of each layer). This means that the mixing enthalpy of the Co and Mo compound always decreases even if all Co atoms react with all Mo atoms in the Co/Mo$_2$C multilayer, or in other words, once Co and Mo compounds are formed and they cannot easily separate.

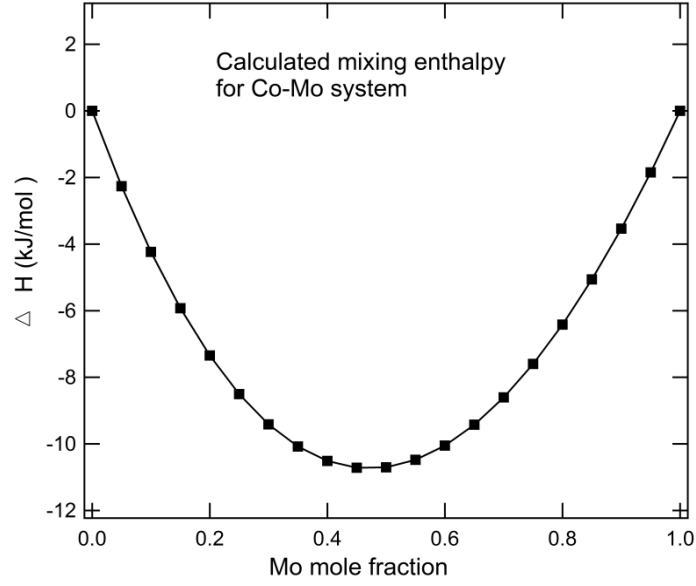

Fig.9 Simulated mixing enthalpy for the Co-Mo system.

The loss of the ferromagnetic Co signal in NMR could be ascribed to a disordered mixture at the interface, such as, $Co_xC_y$. Moreover, mixing enthalpy calculations in Ref. [44] indicate that the Co-C system separate easily into two phases. This is testified by the NMR results of the reference Co/C multilayers. From the chemical state of C atoms and the increase of ferromagnetic Co signal above 300°C, we should pay attention to the change of enthalpy formation for Mo-C mixing system.

For the Mo-C system, the enthalpy change can be expressed as:

$$\Delta H = \frac{2Pc_C f_{Mo}^C \left(c_C V_C^{2/3} + c_{Mo} V_{Mo}^{2/3}\right)\left[-(\phi_C - \phi_{Mo})^2 + \frac{Q}{P}\left(n_{ws}^{C\ 1/3} - n_{ws}^{Mo\ 1/3}\right)^2 - \frac{R}{P}\right]}{n_{ws}^{Mo\ -1/3} + n_{ws}^{C\ -1/3}} + c_C[\Delta H(C_{element} \rightarrow C_{metal})] \quad (3)$$

$$c_{Mo}^S = c_{Mo} V_{Mo}^{2/3} \left(c_{Mo} V_{Mo}^{2/3} + c_C V_C^{2/3}\right)$$

$$c_C^S = c_C V_C^{2/3} \left(c_{Mo} V_{Mo}^{2/3} + c_C V_C^{2/3}\right)$$

P=14.1,    Q/P=9.4 $V^2$/(d.u.)$^{2/3}$    R/P=2.1 $V^2$

Here $\Delta H(C_{element} \rightarrow C_{metal})$ stands for the transformation enthalpy of C into a metallic state and the third term in the equation (3) represents a hybridization contribution and favors a tendency of phase separation. For carbon, the corresponding transformation energy equals the heat of fusion, 100 kJ/mol. To improve the agreement between predicted and experimental values, the transformation energy has been changed to 180 kJ/mol [45].

As mentioned above, we need to consider the volume change of the constituents C and Mo that occurs due to charge transfer when the alloy is formed. The relation is expressed as:

$$V_{C\,alloy}^{2/3} = V_{C\,pure}^{2/3} \cdot \left[1 + a f_{Mo}^{C}(\phi_C - \phi_{Mo})\right] \quad (4)$$

where $a$ is a constant equal to 0.04 for carbon.

The calculated mixing enthalpy of the Mo-C system is presented in Fig.10. We can see that the value of formation enthalpy is negative at C mole fraction below 0.45 and after increases sharply. In the Co/Mo$_2$C stack, the atomic fraction of Mo to C in the Mo$_2$C layer is designed to be 2. That is to say, the Mo$_2$C compound shows small possibility of phase separation. Moreover the thermal stability of Mo$_2$C has been already demonstrated [16,46]. Regarding the results mentioned above, we infer that there is a possibility of excess C atoms existing in the multilayer. Actually, this is demonstrated by the x-ray photoemission spectroscopy results of thin film Mo$_2$C, which gives the atomic ratio of 0.5 between Mo and C atoms.

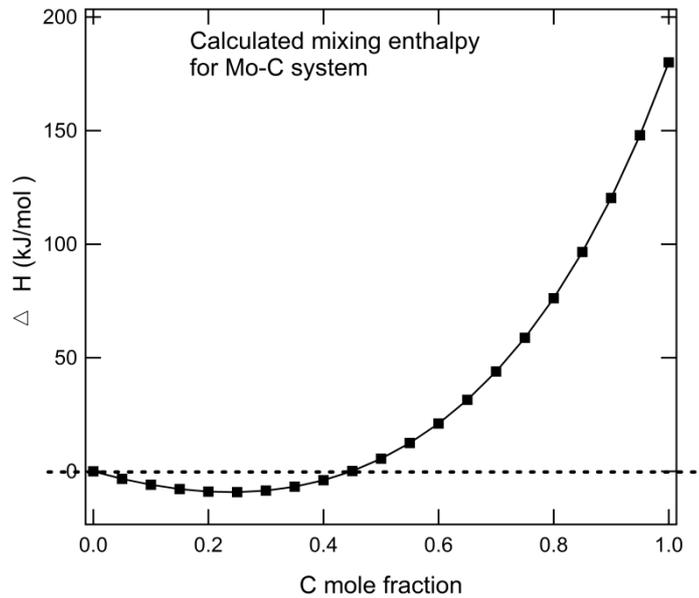

Fig.10 Simulated mixing enthalpy for the Mo-C system.

Based on all the described results, we conclude that the Co atoms in the Co layers are mixing with C atoms for as-deposited and 200°C annealed Co/Mo$_2$C samples. After annealing at 300°C and above, Co and C atoms demix from their mixed regions. However the direction of diffusion may not follow the growth direction of thin layers but other directions, such as a direction parallel to the surface. This means that separated Co and C atoms do not necessarily go back to their original layers. This phenomenon explains the decrease of the reflectivity of the multilayer after annealing [24].

## 5. Conclusion

The interfacial and structural properties of Co/Mo$_2$C multilayers with annealing were characterized by the XES, NMR, ToF-SIMS, XRD and TEM. The Co/Mo$_2$C multilayers show a periodic structure, the Co and Mo$_2$C layers in the stack being mainly amorphous. The formation of oxide layers at both air/stack and stack/substrate interfaces are observed by the ToF-SIMS. According to TEM results, some crystallites are present in the as-deposited samples. The C K$\alpha$ emission spectra of the Co/Mo$_2$C multilayers annealed at 400°C and above show a slight broadening with respect to that of the as-deposited sample, suggesting a slightly change of the carbon chemical state. The NMR spectra demonstrate a demixing phenomenon between Co and C in the multilayer stack after annealing. Comparing the NMR spectroscopy results of the Co/Mo$_2$C multilayers with these of the Co/C reference multilayers, we find that Co and C mix together after deposition and then, after annealing at 300°C and above, Co and C demix from their mixed regions. The Miedema's model is used to calculate the mixing enthalpy and accounts for the phenomenon of phase separation of the Co-C system.

**Acknowledgments**

This work is done under the framework of the international ANR-NSFC COBMUL project (ANR #10-INTB-902-01 and NSFC #11061130549). We would like to thank Imène Estève for preparing FIB sample in IMPMC in Paris.